\documentclass[aps,prb,twocolumn,sd,amsmath,amssymb,final,superscriptaddress]{revtex4-1}
\usepackage{dcolumn}% Align table columns on decimal point
\usepackage{graphicx}% Include figure files
\usepackage{epstopdf}
\usepackage{amsmath}
\usepackage[bookmarks=false,pdfstartview=FitH]{hyperref}
\usepackage{color}
\usepackage{booktabs}
\usepackage[all]{hypcap}

\hypersetup{pdftex,
  breaklinks=true,  % so long urls are correctly broken across lines
  colorlinks=true,
  urlcolor=blue,
  linkcolor=blue,
  citecolor=blue,
  }

\begin{document}

%\preprint{}

\title{Origin of the anomalous semiconducting behaviour in dense lithium}


\author{Mauricio A. Flores}
\email[]{mauricio.flores@ug.uhile.cl}
%\thanks{}
\affiliation{Facultad de Ingenier\'ia y Tecnolog\'ia, Universidad San Sebasti\'an, Bellavista 7, Santiago, 8420524, Chile.}

\author{Gonzalo Guti\'errez}

\affiliation{Departamento de F\'isica, Facultad de Ciencias, Universidad de Chile, Casilla 653, Santiago, Chile.}


\begin{abstract}
Experimentally, it is known that lithium undergoes a metal to semiconductor transition at about 80 GPA and a reentrant semiconductor to metal transition near 120 GPA. This unusual behaviour has been attributed to the formation of high-pressure electrides in the Li-\textit{Aba}2 phase. Using the accurate wave function based quantum Monte Carlo (DMC) method, we show that the valence charge distribution of the Li-\textit{Aba}2 phase is incompatible with an insulating or semiconducting ground state. At DMC level, the most stable phase at 100 GPA is an orthorhombic oP24 structure with Pbca symmetry whose valence charge density shows an electride paired distribution, in correspondence with the theoretical predictions of Neaton and Ashcroft [Nature 00, 141 (1999)]. Here, we propose the electride pairing in the oP24-(Pbca) phase as the origin of the semiconducting behaviour observed in diamond anvil cell experiments.
\end{abstract}

\date{\today}

%\pacs{}
%\keywords{}

\maketitle

Under high pressure lithium exhibits an exotic and often counterintuitive chemistry. It shows an anomalous melting curve, \cite{Lazicki10,Guillaume11,Schaeffer12} isotope effects,\cite{Struzhkin02,Schaeffer15,Ackland17} and a Fermi surface that becomes progressively anisotropic with increasing pressure. \cite{Boettger85,Rodriguez06} In 1999, Neaton and Ashcroft \cite{Neaton99} predicted that at high pressures lithium should undergo a metal-to-insulator (or semiconductor) transition to a paired structure, similar to a condensed molecular solid as observed in dense hydrogen. \cite{Johnson00} This theoretical prediction was confirmed experimentally ten years later by Matsuoka and Shimizu \cite{Matsuoka09} who observed a metal to semiconductor transition at about 80 GPA, \cite{Matsuoka09} and four years later, a reentrant semiconductor to metal transition near 120 GPA. \cite{Matsuoka14}

\begin{figure}[h]
\vspace{0.2cm}
 \includegraphics[width=6.2cm]{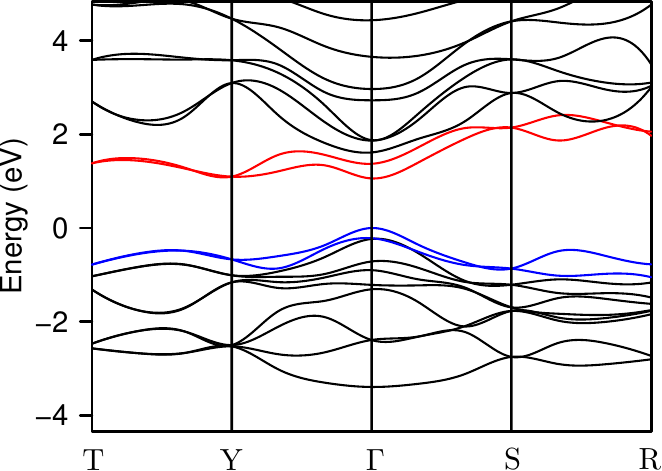}\vspace{0.2cm}
  \caption{\textit{QSGW} quasiparticle band structure of the Li-\emph{Aba}2 phase at 60 GPA.}\label{fig:1}
\end{figure}

Several theoretical studies based on density-functional theory (DFT) have proposed the (\textit{Aba}2, oC40) phase (hereinafter referred to as \textit{Aba}2) as the most plausible candidate for the semiconducting regime. \cite{Lv11,Marques11,Gorelli12} In this phase the valence electrons separate from all atoms and tend to occupy interstitial regions, leading to the formation of high-pressure electrides. \cite{Singh93,Dye09,Lee13} Both DFT and many-body  \emph{GW} \cite{Hedin65} calculations predict the existence of a semiconducting band gap in the Li-\textit{Aba}2 phase.\cite{Marques11,Miao17,Yu18} However, it has been pointed out that some of the interstitial sites are singly occupied.\cite{Miao17} Thus, in principle, this phase should be metallic. However, this fundamental contradiction may be alleviated if two neighboring interstitial quantum orbitals actually form covalent bonds (i.e., a \emph{quasimolecule}) and, therefore, a gap opens up between their bonding and antibonding states, as recently proposed by Miao and co-workers. \cite{Miao17}

\begin{figure}[h]
 \centering
 \includegraphics[height=3.8cm]{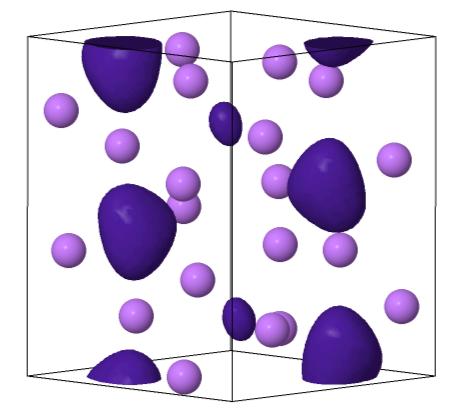}\hspace{-0.1cm}
 \includegraphics[height=3.8cm]{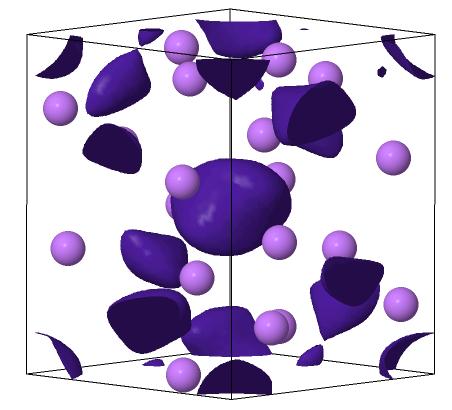}
  \caption{DFT charge density isosurfaces ($\rho = 0.003 \hspace{0.1cm} e/\text{bohr}^3$) of (Left) the valence-band maximum, and (Right) the conduction-band minimum of the Li-Aba2 phase at 60 GPA. The violet spheres represent Li nuclei.}\label{fig:2}
\end{figure}

\begin{figure*}
 \centering
 (a)\hspace{0.2cm}\includegraphics[width=6.5cm]{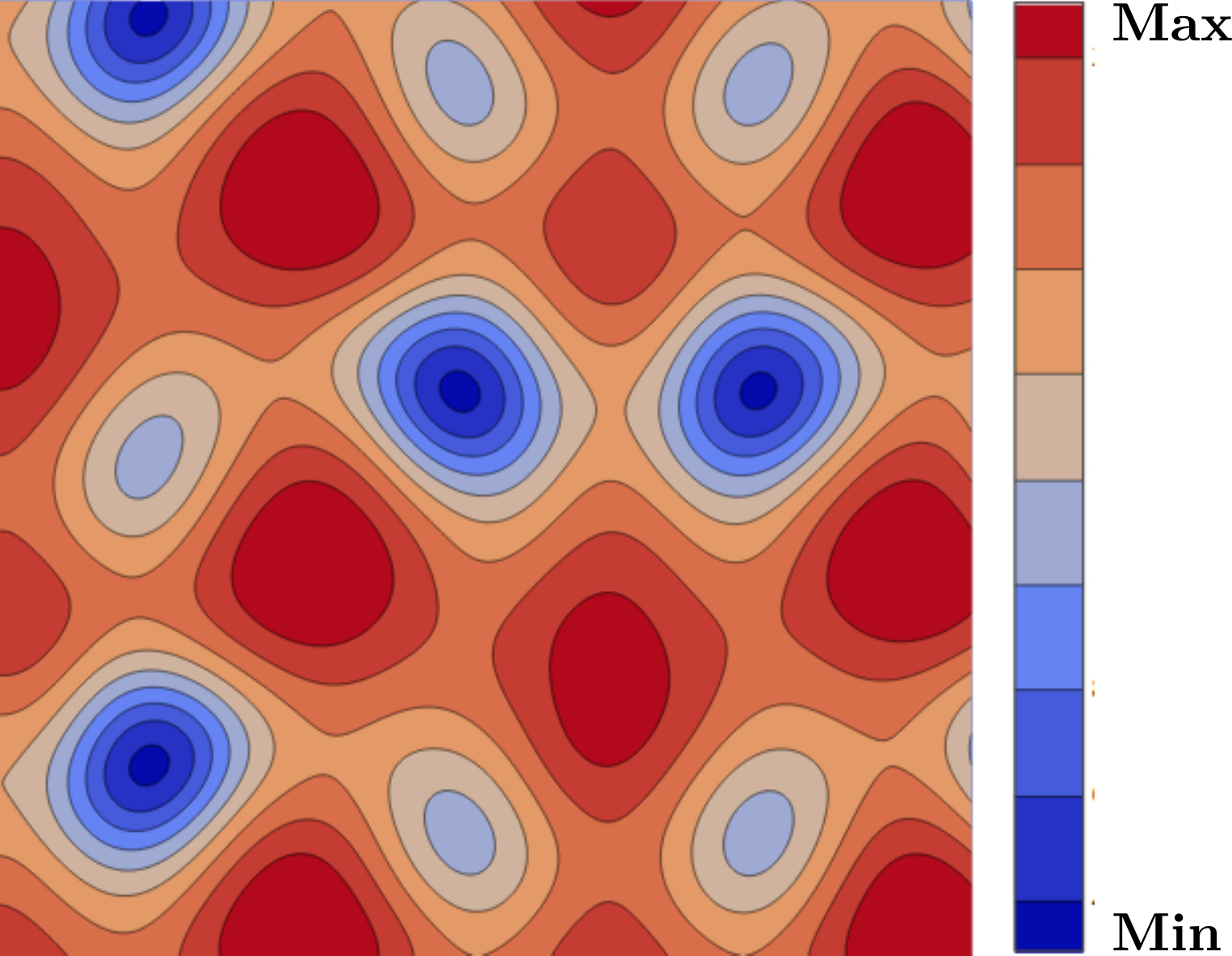}
 \hspace{1cm}
 (b)\hspace{0.2cm}\includegraphics[width=6.5cm]{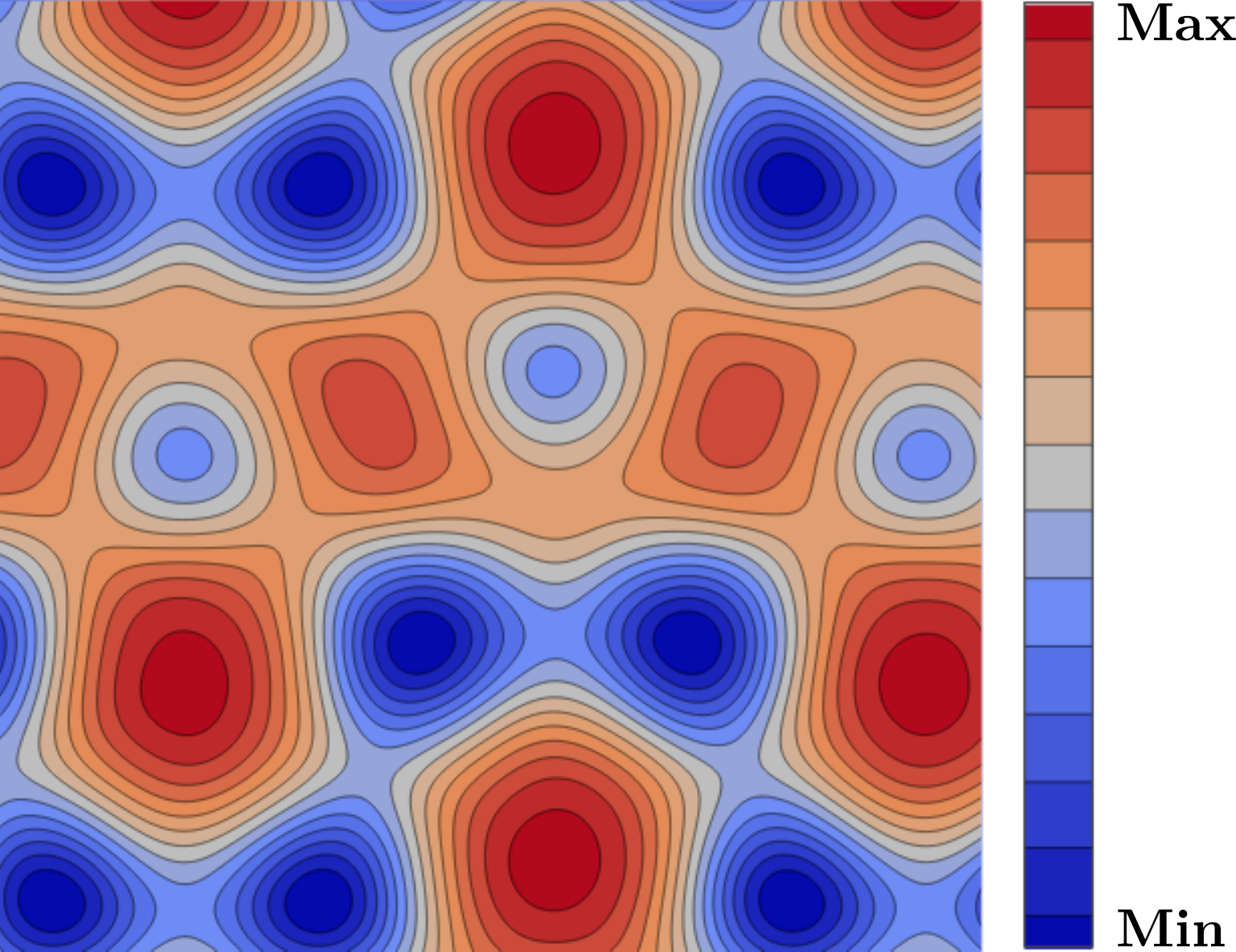}
  \caption{DMC valence charge density of (a) Li-Aba2 at 60 GPA and (b) Li-cI16 phase at 60 GPA, plotted in the Miller index plane (100).}\label{fig:3}
\end{figure*}

In this Letter, we investigate the origin of the semiconducting behaviour experimentally observed in dense lithium. We study the high-pressure Li-\textit{Aba}2 structure, which has been identified previously as the best candidate for the semiconducting phase of Li, by combining DFT and many-body \emph{GW} calculations with the highly accurate diffusion quantum Monte Carlo (DMC) method. The topology of the valence electron distribution of the Li-\textit{Aba}2 phase reveals the existence of partially covalent multi-center bonds centered on interstitial electrides. Moreover, our DMC results indicate that the most stable structure at 100 GPA is an orthorhombic oP24 structure with Pbca symmetry, whose valence charge distribution shows the existence of electride-electride pairs or \emph{quasimolecules} localized at interstitial sites, being consistent with an
insulating or semiconducting ground state as described by Kohn. \cite{Kohn64} 

 In in 2011, the high-pressure Li-\textit{Aba}2 structure was identified in two computational studies. \cite{Marques11,Lv11} Marqu\'es \emph{et al.} \cite{Marques11} and Lv \emph{et al.} \cite{Lv11} reported a Li-\textit{Aba}2 structure as the lowest-enthalpy phase in the pressure range of 67-91.3 GPA and 60-80 GPA, respectively. These results are in good agreement with x-ray diffraction data, \cite{Guillaume11} which suggests a C-face centred orthorhombic structure having 40 atoms per unit cell (oC40) as the most stable phase between $\sim$70 and $\sim$100 GPA. Guillaume \emph{et al.}\cite{Guillaume11} reported the existence of a transition from this phase to an orthorhombic structure near 100 GPA at 77 K, which has 24 atoms per unit cell and seems to be stable to at least 130 GPA. Moreover, as Li-\textit{Aba}2 is the only known competitive structure showing a semiconducting gap in DFT calculations, it has been systematically associated in the literature with the semiconducting phase experimentally observed in the pressure range of 80-120 GPA. Naumov and Hemley,\cite{Naumov15} assuming a weakly correlated system, suggested that the opening a band gap in the Li-\textit{Aba}2 phase results from a strong $s$-$p$ orbital mixing near the Fermi level. Miao \emph{et al.}  \cite{Miao17} suggested that its semiconducting nature could be due to the formation of bonding and antibonding states (\textit{quasimolecules}) between two neighbouring interstitial electrides.

   \hspace{0.35cm}We first investigate the electronic structure of the \textit{Aba}2 phase at 60 GPA within the quasiparticle self-consistent \textit{GW} approximation (\emph{QSGW}), which is known to provide an accurate description of the electronic structure of solids. \cite{Shishkin07,Van06,Flores17} Our calculations were performed using the ABINIT code, \cite{Gonze2009} with full details provided in the Supplementary Information. Figure \hyperref[fig:1]{1} shows the quasiparticle band structure of the Li-\textit{Aba}2 phase at 60 GPA. We found the existence of a direct gap of 1 eV at the $\Gamma$ point, thereby confirming the semiconducting character of this phase as predicted by previous DFT calculations.
      
   In addition, we have found that both band edges are derived from interstitial electrides. In Figure \hyperref[fig:1]{1}, the blue bands in the valence-band are mainly derived from electrides with a predominant $s$-like character, whereas the red bands at the bottom of the conduction-band are derived from electrides that are a mixture of $s$- and $p$-like states. The charge density isosurfaces corresponding to the valence-band maximum (VBM) and the conduction-band minimum (CBM) are plotted in Figure \hyperref[fig:2]{2}. It is worth to mention that the electrides that form the band edges could be seen as periodic arrays of quantum dots, \cite{Dye09} whose wave functions strongly overlap due to their high concentration and close proximity, \cite{Mott56,Winkler11,Flores18} giving rise to delocalized bands in reciprocal space.

\begin{figure*}
 \centering
 \includegraphics[height=7.2cm]{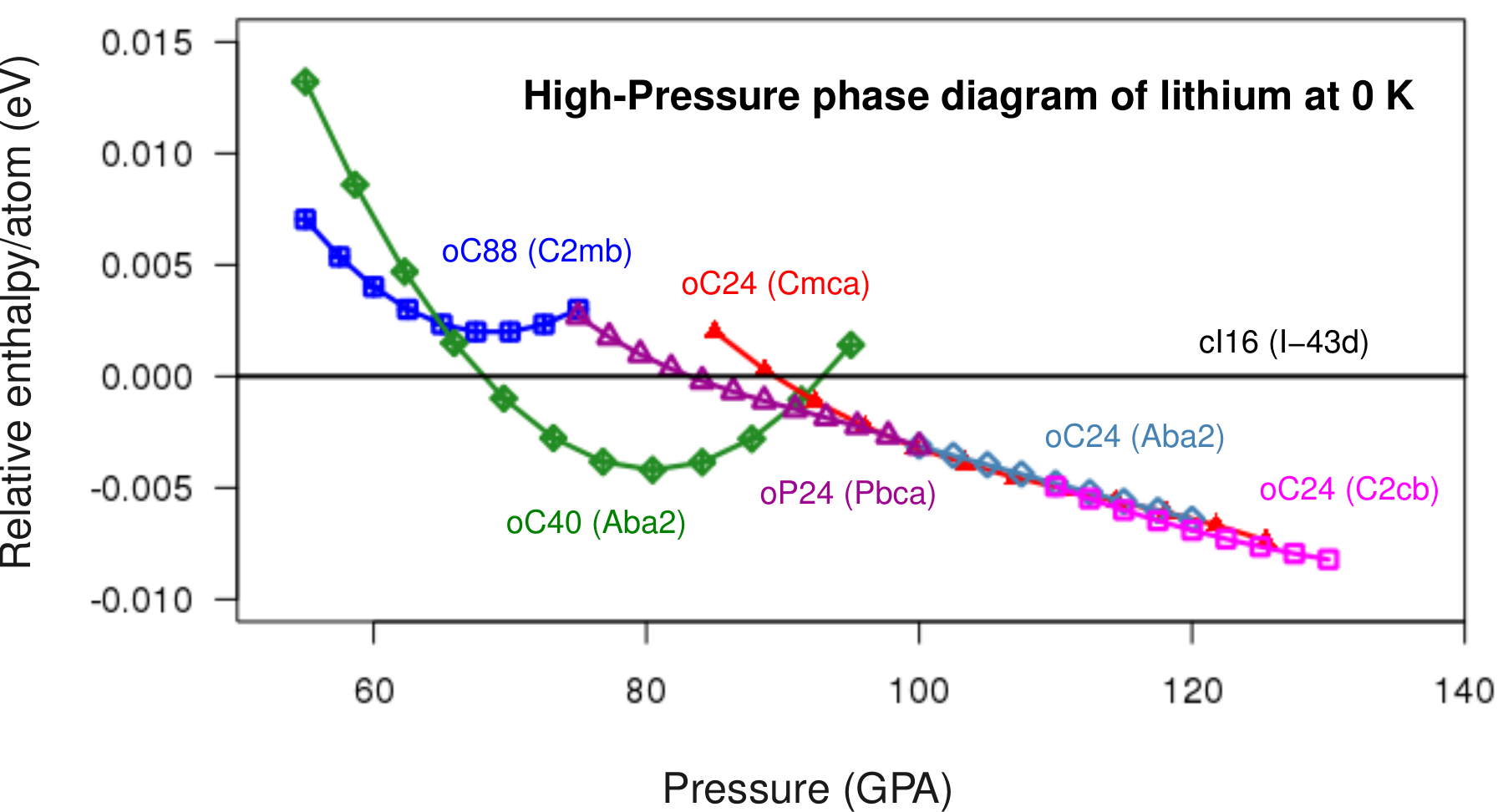}
 \caption{Phase stability diagram of Li as a function of pressure for the range of 50 to 140 GPA obtained by DFT calculations using the Perdew-Burke-Ernzerhof (PBE) \cite{Perdew96} exchange-correlation functional, where enthalpy is given by $H = E + PV$. Resistivity measurements performed by Matsuoka \emph{et al.} \cite{Matsuoka09,Matsuoka14} indicate that Li exhibits a semiconducting behaviour between 80 and 120 GPA. At 100 GPA, DMC lowers the enthalpy of the oP24-Pbca structure by 9.77 meV/atom relative to the oC24-Cmca phase.}\label{fig:4}
\end{figure*}

 The semiconducting nature of Li-\textit{Aba}2 is in disagreement with experimental observations, which indicate that Li is metallic in the range of 70-80 GPA. \cite{Matsuoka09,Matsuoka14} However, the Pauli exclusion principle that prevents short-range interactions may play a crucial role in determining the electronic properties of the Li-\emph{Aba}2 phase. If the Pauli repulsion is strong enough, the valence charge density could be rejected from the position of the $s$-like electrides that form the top of the valence-band (blue bands in Figure \hyperref[fig:1]{1}) and be partially transferred to the high energy electrides of mixed $s$- and $p$-like character that form the bottom of the conduction-band (red bands in Figure \hyperref[fig:1]{1}), thus avoiding the overlap between the negatively charged $s$-like electrides and the core $1s$ electrons of the neighboring Li nuclei. Moreover, if the electrides that form the top of the valence-band are primarily 1\emph{s} in character (as is the case of sodium near 200 GPA) \cite{Miao15} the possibility of a charge transfer between the two classes of electrides should be strongly increased. To investigate this hypothesis, we used the accurate wave function based diffusion quantum Monte Carlo method \cite{Foulkes01,Austin12} to obtain the valence electron charge distribution of the Li-\emph{Aba}2 phase at 60 GPA. The DMC method is in principle an exact technique to solve the imaginary time dependent Schr\"{o}dinger equation. It has been successfully applied in the study of the high-pressure phase diagram of solid molecular hydrogen,\cite{Drummond15} the high-pressure properties of silica,\cite{Driver10} and the stability of square 2D ice under pressure. \cite{Chen16}  We have used a trial wave function of the Slater-Jastow form, \cite{Drummond04} in which a determinant of single-particle orbitals obtained from a DFT calculation is multiplied by a Jastrow correlation factor. In order to make the wave function depend explicitly on the particle separations, we used a three-term Jastrow factor that includes an electron term, an electron-nucleus term, and an electron-electron-nucleus term. In Supplementary Note 3, we compare the accuracy of DFT and DMC for the description of the dissociation curve of the H$_2$ molecule, the simplest case of a correlated many-electron system. 
 
According to Kohn, \cite{Kohn64} the insulating behavior is a strict consequence of the localization of the electronic wave function in the many-particle configuration space. This characterization includes conventional insulators, in which the ground state is isolated from the exited states by a finite energy gap, as well as Mott insulators, \cite{Mott49,Mott56,Mott61} which are usually described as metals by one-electron band theory. Figure \hyperref[fig:3]{3(a)} shows the valence charge distribution of the Li-\emph{Aba}2 phase at 60 GPA obtained from quantum Monte Carlo calculations. The DMC result, in which correlations were explicitly included through a Jastrow correlation factor, is consistent with our previous analysis. The valence density is mostly delocalized and exhibits several local maxima centered at interstitial positions. We observe a large degree of delocalization between the basins and wide regions in which the electron distribution is relatively flat, being consistent with metallic behavior. \cite{Silvi00} Moreover, the topology of the electron distribution indicates a multicenter bonding situation similar to that previously reported for the metallic Li-cI16 phase. \cite{Hanfland00} In addition, DMC shows that the valence electrons percolate  through the space forming a three-dimensional connected network, as shown in Supplementary Figure 2(b). These findings seem to be incompatible with the formation of isolated covalently bonded \emph{quasimolecules} in the Li-\emph{Aba}2 phase, as recently proposed in Ref. [\onlinecite{Miao17}].  

It is illustrative to compare the DMC valence charge distributions of Li-\emph{Aba}2 and Li-cI16, as the latter phase is well-known to be metallic and is also characterized by the formation of high-pressure electrides. \cite{Alonso06,Yu18} The calculated valence charge distribution of the metallic Li-cI16 phase is shown in Figure \hyperref[fig:3]{3(b)}. The more delocalized character in the Li-\emph{Aba}2 phase (Figure \hyperref[fig:3]{3(a)}) is clear. Thus, the appearance of a finite energy gap seems to be incompatible with the Kohn's description of the insulating state.

\begin{figure*}
 \centering
 (a)\hspace{0.2cm}\includegraphics[width=8.1cm]{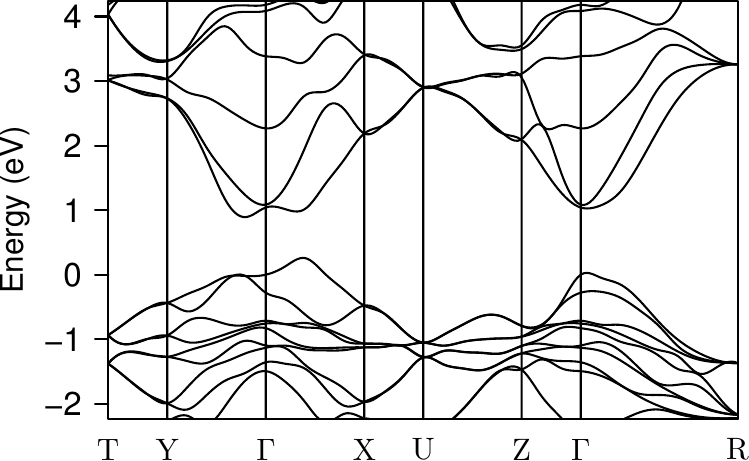}
 \hspace{1cm}
 (b)\hspace{0.3cm}\includegraphics[width=6.5cm]{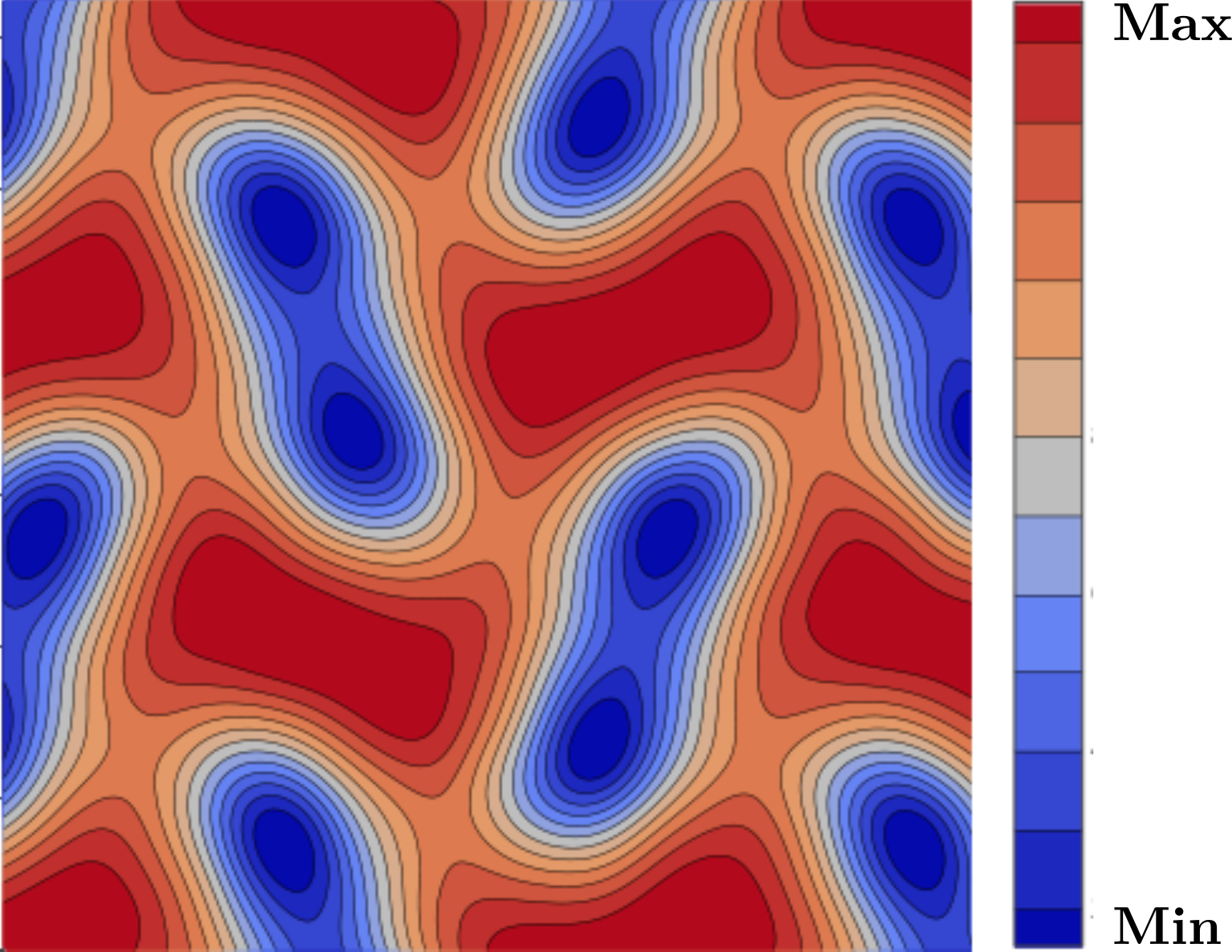}
  \caption{(a) Electronic structure of Li-oP24 at 100 GPA. A scissors operator, consisting of a rigid shift to the conduction bands so as to recover the DMC quasiparticle band gap, was applied to the DFT-PBE band structure. (b) DMC valence charge density of Li-oP24 at 100 GPA plotted in the Miller index plane (002). Lithium ions are paired with a dimer distance of 1.55 \AA.}\label{fig:5}
\end{figure*}

Finally, we performed DFT calculations to obtain the static (i.e., neglecting the zero-point motion) low temperature phase diagram of Li between 50 and 140 GPA, including the most competitive crystal structures previously proposed in the literature.\cite{Rousseau05,Pickard09,Marques11,Gorelli12} Our results are shown in Figure \hyperref[fig:4]{4}. We found that the Li-\emph{Aba}2 structure is the most stable phase in the range of 68.5$-$90.5 GPA, in agreement with previous theoretical works. \cite{Marques11,Lv11,Gorelli12} Furthermore, we chose 100 GPA as a representative pressure for the experimentally observed semiconducting regime. At this pressure, we identify three distinct structures nearly degenerate in enthalpy: a local three-coordinated oC24 (\emph{Cmca}) structure (proposed by Rousseau \emph{et al.}),\cite{Rousseau05} an orthorhombic oP24 (\emph{Pbca}) structure, and an oC24 structure with \emph{Aba}2 symmetry. All of these are electride phases, showing valence charge localization at interstitial positions. 

We performed DMC calculations to compare their relative enthalpies at a higher level of theory and found that Li-oP24 (\emph{Pbca}) is the most stable phase at 100 GPA. The second most stable phase, higher 9.77 meV/atom in enthalpy, is Li-oC24 (\emph{Cmca}) which is closely followed by Li-oC24 (\emph{Aba2}), just 1.79 meV/atom higher in enthalpy. We should note that the enthalpy of insulating phases is expected to be lowered at DMC level with respect to metallic phases. \cite{Johnson00} This is due to a downshift in the absolute position of the valence-band that lowers the total energy. In metallic phases, there are no such downshifts and DFT is expected to be sufficiently accurate. We can obtain the correction to the valence-band of the Li-oC24 phase by comparing the ionization potential (IP) obtained by PBE and DMC calculations. The IP can be approximated as, IP $=E_0 - E_+$, where $E_0$ is the total energy of the ground state and $E_+$ is the total energy of the positively charged state. \cite{Ertekin13} We found a correction of $-0.27$ eV for the valence-band. Similarly, the correction to the conduction-band can be approximated by computing the electron affinity (EA) given by, EA $=E_- - E_0$, where $E_-$ is the total energy of the negatively charged state. We found a correction of $0.32$ eV for the conduction-band by comparing the PBE and DMC results.

Figure \hyperref[fig:5]{5(a)} shows the electronic band structure of the Li-oP24 phase at 100 GPA, in which DMC corrections of $0.32$ eV and $-0.27$ eV were applied to the conduction-band and the valence-band, respectively. Figure \hyperref[fig:5]{5(b)} shows the DMC valence charge distribution plotted in the Miller index plane (002). We find that the distribution is mostly localized showing electride-electride pairs or \emph{quasimolecules} centered at interstitial sites. These findings are in good agreement with the theoretical predictions of Neaton and Ashcroft, \cite{Neaton99} suggesting that electride pairing in the Li-oP24 phase could be the origin of the semiconducting behaviour observed in diamond anvil cell experiments. \cite{Matsuoka09,Matsuoka14} Moreover, the calculated lattice parameters at 115 GPA are $a=4.249$ \AA, $b=4.236$ \AA, and $c=7.581$ \AA, in good agreement with the experimentally found values of $a=4.213$ \AA, $b=4.205$ \AA, and $c=7.482$ \AA. \cite{Guillaume11} 

In summary, our calculations exclude the formation of high-pressure electrides in the Li-\emph{Aba}2 phase as the origin of the semiconducting behavior experimentally observed in compressed lithium. Our DMC results, in which electron correlations are explicitly taken into account, show that the most stable crystal structure at 100 GPA is a paired orthorhombic oP24 structure with Pbca symmetry. Thus, we propose the electride-electride pairing in this phase as the origin of the semiconducting behaviour experimentally observed in the range of 80-120 GPA.\vspace{0.2cm}\\

\begin{acknowledgments}
 Powered@NLHPC: This research was partially supported by the supercomputing infrastructure of the NLHPC (ECM-02). G.G. acknowledges
 the finantial support of Fondo Nacional de Investigaciones Cient\'ificas y Tecnol\'ogicas (FONDECYT, Chile) under Grant No. 1171127.
\end{acknowledgments}

\bibliographystyle{apsrev4-1}
\bibliography{Lithium}

\end{document}